\documentclass[10pt]{article}
\usepackage{mathrsfs}
\usepackage{amsmath,amsfonts,amssymb,mathtools}
\usepackage{cite}
\usepackage{dsfont}
\usepackage{graphicx}
\usepackage{hyperref}
\usepackage{verbatim}
\usepackage{slashed}
\usepackage[all]{xy}
\usepackage{xcolor}
\usepackage{subfig}
\usepackage{tikz}
\usetikzlibrary{shapes.geometric,arrows}
\usepackage{relsize}
\usepackage{caption}
\usepackage{float}
\usetikzlibrary{positioning}
\usepackage{geometry}
\usepackage[T1]{fontenc}
\usepackage{lmodern}
\usepackage{stackengine} 
 
\usepackage[utf8]{inputenc}

\hypersetup{
  colorlinks,
  citecolor=violet,
  linkcolor=blue,
  urlcolor=blue}

\csname @addtoreset\endcsname{equation}{section}
\DeclareMathAlphabet\mathbfcal{OMS}{cmsy}{b}{n}

\newcommand{\beq}{\begin{equation}}
\newcommand{\eeq}{\end{equation}}
\newcommand{\bea}{\begin{eqnarray}}
\newcommand{\eea}{\end{eqnarray}}
\newcommand{\ba}{\begin{array}}
\newcommand{\ea}{\end{array}}
\newcommand{\bit}{\begin{itemize}}
\newcommand{\eit}{\end{itemize}}
\newcommand{\nn}{\nonumber}

\newcommand{\mezzo}{\frac{1}{2}}
\newcommand{\complesso}{{\ \hbox{{\rm I}\kern-.6em\hbox{\bf C}}}}
\newcommand{\reale}{{\hbox{{\rm I}\kern-.2em\hbox{\rm R}}}}
\newcommand{\uno}{ \,  \raisebox{+0.14em}{{\hbox{{\rm \scriptsize ]}} \raisebox{-0.2em}{\kern-.8em\hbox{1}}}} \, }  

\newcommand{\p}{\partial}

\renewcommand{\d}{\delta}
\newcommand{\D}{\Delta}
\newcommand{\e}{\epsilon}

\renewcommand{\l}{\lambda}
\renewcommand{\L}{\Lambda}

\newcommand{\m}{\mu}

\newcommand{\n}{\nu}

\newcommand{\s}{\sigma}

\renewcommand{\S}{\Sigma}

\begin{document}


\begin{titlepage}

\vspace{0.3cm}

\begin{flushright}
$LIFT$--12-5.25
\end{flushright}

\vspace{1.0cm}

\begin{center}
\renewcommand{\thefootnote}{\fnsymbol{footnote}}
\vskip 9mm  
{\Huge \bf Static hairy black hole
\vskip 10mm
   in 4D General Relativity}
\vskip 37mm
{\large {Marco Astorino$^{a}$\footnote{marco.astorino@gmail.com}
}}\\

\renewcommand{\thefootnote}{\arabic{footnote}}
\setcounter{footnote}{0}
\vskip 8mm
\vspace{0.2 cm}
{\small \textit{$^{a}$Laboratorio Italiano di Fisica Teorica (LIFT),  \\
Via Archimede 20, I-20129 Milano, Italy}\\
} \vspace{0.2 cm}
%

\end{center}

\vspace{2.7cm}

\begin{center}
{\bf Abstract}
\vspace{1mm}
\end{center}
{In four-dimensional vacuum general relativity the only known static, exact and analytical black hole solution is given by the Schwarzschild spacetime. In this paper this renowned metric is generalised by adding another integrating constant, a hair that  switches the metric from the Petrov type $D$ to the type $I$. This new parameter represents the intensity of an external gravitational field, which can be considered the hyperbolic generalisation of the Witten's bubble of nothing. No curvature or conical singularities are present outside the event horizon.
The no hair arguments are circumvented because the metric is not asymptotically flat, and neither the black hole is spherical. \\
The gravitational hair continuously deforms the Schwarzschild geometry: the horizon becomes oblate, while its area is reduced. Conserved charges and thermodynamic properties of the black hole are studied.}

\end{titlepage}

\addtocounter{page}{1}

\newpage


\section{Introduction}
\label{sec:introduction}

In four dimensional general relativity, described by the Einstein field equations $R_{\m\n}=0$, the only black hole static exact solution known in the literature is the Schwarzschild solution from 1916 \cite{schwarzschild}.
We are referring to analytical metrics completely regular in the domain of outer communication\footnote{A class of Schwarzschild generalisations, called {\it distorted black holes} representing black holes in an external gravitational field have been proposed, these could work well as local models, not far from the event horizon, since at spatial infinity the scalar curvature invariant are unbounded, see \cite{chandrasekhar}. However because of this feature these distorted black holes are not always considered completely regular.}. There exists a diagonal generalisation of the renowned spherical symmetric black hole, found by Levi-Civita in 1918 \cite{levi-civita-c}, known as the C-metric, but since it is accelerating it cannot be considered completely static. In fact the accelerating horizon defines some non-static regions outside the event horizon. Similarly the Schwarzschild black hole inside the bubble of nothing \cite{bubble}, although it is diagonal, the complete metric can hardly be considered static because the black hole is immersed into an expanding bubble. It is worth to mention that a recent static black hole generalisation of the Schwarzschild black hole has been built in \cite{swirling}, however, even though the black hole horizon is not rotating,  it is embedded in a swirly rotating background, so the full solution is actually stationary.\\
In the last decades a number of no hair theorems have been produced to assure the unicity of the Schwarzschild solution, the most famous being\cite{jebsen}, \cite{birkhoff}, \cite{israel}. In any case they mainly rely on the asymptotic flatness or spherical symmetry assumptions.\\
In this letter we get rid of these two strong hypotheses to pursue an extension of the Schwarzschild black hole. Our physical motivation stems from the fact that black holes usually are not completely isolated; indeed those effectively detected and studied in nature are, in general, surrounded by matter. This matter is not necessarily distributed in a perfectly isotropic symmetrical way, but more often the external matter is spread in accretion disks. Hence it can be beneficial also to relax the spherical symmetric assumption, in order to obtain more or less pronounced deformation of the Schwarzschild black hole. \\
The metric we consider can be generated, starting from the black hole in the external Bertotti-Robinson magnetic field presented in [16], by the symmetry transformations of the Ernst equations, as described in \cite{bh-brbm}. The Ernst equations are a reformulation of the Einstein field equations for axisymmetric and stationary spacetime \cite{ernst1}, which admits a prompt use of the equations symmetry. We will describe the geometrical and physical properties of the novel black hole, focusing mainly on studying its thermodynamics and in identifying its gravitational background.\\

\section{Static, type I, hairy black hole solution in vacuum}
\label{sec:solution}

\subsection{Geometrical and physical properties}
\label{sec:geometry}

\begin{figure}[h!]
\captionsetup[subfigure]{labelformat=empty}
\centering
\hspace{-0.5cm}
\subfloat[\hspace{0.1cm} $({\bf a})$ \ $m=1$, $B=0$]{{\includegraphics[scale=0.4]{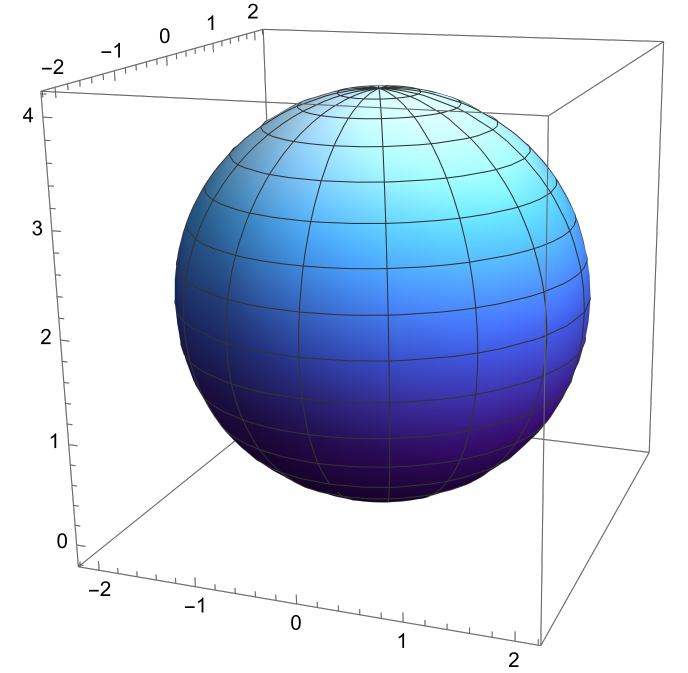}}}%
\subfloat[\hspace{0.1cm} $({\bf b})$ \ $m=1$, $B=0.4$]{{ \hspace{0.5cm} \includegraphics[scale=0.4]{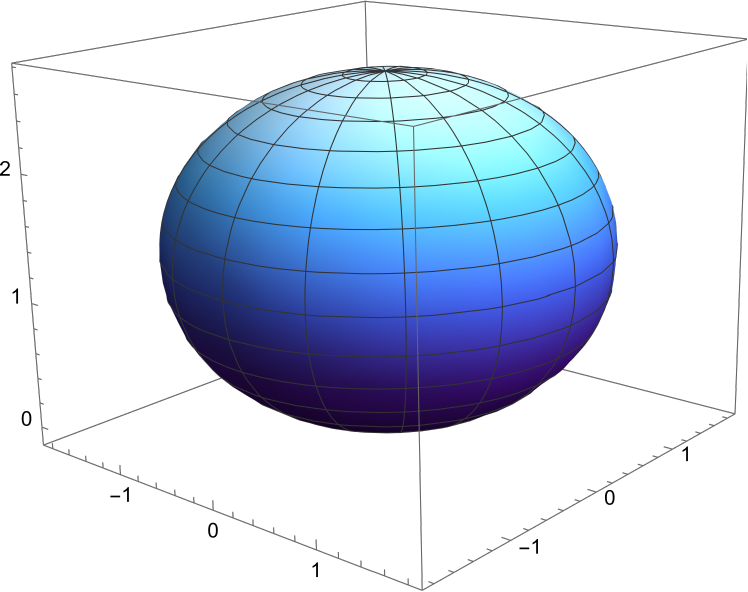}}}%
\subfloat[\hspace{0.3cm} $({\bf c})$ \ $m=1$, $B=0.5$]{{\hspace{0.1cm}
\includegraphics[scale=0.4]{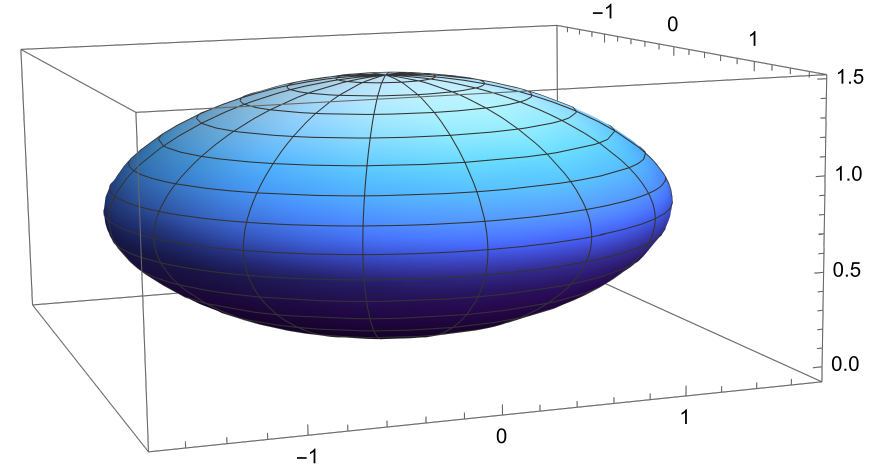}}}
\caption{\small Embedding in Euclidean three-dimensional space $\mathbb{E}^3$ of the black hole's event horizon distorted by the presence of an external gravitational field, for different values of the parameter $B$, while keeping $m=1$. When $B=0$ we have the usual spherical Schwarzschild horizon, as in picture (\ref{fig:picture-horizons}.a).}%
\label{fig:picture-horizons}
\end{figure}

The new metric we present in this article is\footnote{A Mathematica notebook containing this solution is provided, as an ancillary file in the arXiv source folder or as supplementary material in the PRD journal.} 
\beq  \label{bh}
         ds^2 = \frac{(1+B^2mr\cos^2\theta +\Omega)^2}{4\Omega^4} \left[- f(r) dt^2 + \frac{dr^2}{f(r)} + \frac{r^2 d\theta^2}{1+B^2m^2\cos^2\theta} \right] + \frac{4 r^2 \sin^2 \theta \, (1+B^2m^2\cos^2\theta) \, \D_\varphi^2 \, d\varphi^2}{(1+B^2mr\cos^2\theta +\Omega)^2} \, ,
\eeq
where 
\bea
        f(r) &=& \left(1-\frac{2m}{r} -B^2m^2\right)(1+B^2r^2) \ . \\
        \Omega(r,\theta) &=& \sqrt{1+B^2r[r+(2m+B^2m^2r-r)\cos^2\theta]} \ .\label{Omega}
\eea
This is a solution of vacuum Einstein's general relativity, that is $R_{\m\n}=0$, which describes a generalisation of the Schwarzschild black hole. In fact, for $B=0$, we recover precisely the Schwarzschild spacetime. The three physical constant parameters, $m, B , \Delta_\varphi $ that appear in the metric (\ref{bh}) are related to the mass of the black hole, to the intensity of the external gravitational field and to possible angular defects on the azimuthal axis. \\
The causal structure is the same of the Schwarzschild-Bertotti-Robinson metric \cite{kerr-bertotti}, because the conformal factor of the first three terms of the metric (related to the $[t,r,\theta]$ coordinates), is always strictly positive and does not diverge, while the remaining parts are identical. Thus the Penrose diagrams are the same as the Schwarzschild-Bertotti-Robinson ones. Regarding the geodesic motion of massive test particles, the metric (\ref{bh})-(\ref{Omega}) shares similarities with the Schwarzschild geometry because, as shown in \cite{kerr-bertotti}, Schwarzschild in the Bertotti-Robinson electromagnetic fields, has formally the same expression for the innermost stable circular orbit with respect to the Schwarzschild black hole. For more details about the geodesic motion see section \ref{sec:geo}.\\
The spacetime (\ref{bh})-(\ref{Omega}) is regular outside the event horizon, because of its symmetry across the equatorial plane (i.e. $\theta \to - \theta$). In fact, to avoid conical singularities on the azimuthal axis, as can be checked eq. (\ref{conicity}), we can set the gauge constant
\beq
            \D_\varphi = \frac{1}{1+B^2m^2} \ .
\eeq
Thus no cosmic strings or struts affect the solution.\\  
The Kretschmann curvature invariant, $R_{\m\n\s\l} R^{\m\n\s\l}$, diverges only for $r=0$, as in the Schwarzschild case, as can be appreciated from the curvature invariant denominator, which is proportional to
\beq
       r^6 \Big\{1+B^2mr\cos^2\theta+\sqrt{1+B^2 r \big[r+(2m-r+B^2m^2r)\cos^2\theta \big]} \Big\}^8 \nn \ .
\eeq
Furthermore, for large radial distances, the Kretschmann scalar invariant remains always a finite function of $\theta$, which can be zero for certain values of the polar angle's coordinate.\\ 
When $B\neq 0$ we get an extension of the Schwarzschild spacetime. Indeed, according to the Petrov classification, the Type of the new solution (computed as explained in \cite{stephani-big-book}, \cite{Griffiths-Podolsky} or \cite{Type-I}) is more general with respect to the $D$-type of the spherical symmetric black hole, that is $I$. Therefore the new solution cannot be a diffeomorphism of the Schwarzschild metric. In any case, it also possesses just an event horizon, but it is located at
\beq \label{horizon}
          r_h = \frac{2m}{1-B^2m^2} \ .
\eeq
Note that, in order to preserve the black hole interpretation and to avoid naked singularities, the radius $r_h$ has to be positive, thus the $B$ parameter must satisfy $|B|<1/m$, while $m$ has to be considered positive\footnote{Another possible black hole interpretation, i.e. for $r_h>0$, would be consider $m<0$ and $B^2m^2>1$, but this parametric sector does not contain a limit to the Schwarzschild black hole, for $B \to 0$.} to be coherent with the Schwarzschild black hole limit (for $B=0$) and not violate viable physical energy conditions. In fact, the positivity of the mass in general relativity is granted unless the dominant energy condition is violated. For the space-time under consideration the the positivity of the mass is determined by the sign of $m$, as can be appreciated from eq. (\ref{mass}).   \\
The main difference with respect to the Schwarzschild black hole is given by the new integration constant $B$, which can be considered a primary hair\footnote{Note that, upon rescaling of the time and radial coordinates and the mass parameter as follows $t \to t/B, \ r \to r/B , m \to m/B$, the $B$ parameter can be regarded only as a constant conformal factor for the metric (\ref{bh}), with $B=1$. The constant conformal factor is $B^{-2}$. This is a typical feature of continuous deformations of the Schwarzschild black hole, take for instance the Schwarzschild-Melvin \cite{ernst-magnetic}, the Reissner-Nordstrom or the Kerr spacetimes. In all these cases it is possible to pick a unit system where the extra integration constant, with respect to the black hole mass, is unitary in the metric, up to an overall constant conformal factor. In particular, considering the above examples, we can set to a fixed value, such as $1$, the intensity of the external Melvin electromagnetic field, of the Coulomb potential or the angular momentum (up to a constant overall conformal factor depending on the extra integrating constant). In practice this fact can be understood as the conformal structure of these Schwarzschild deformations, thus its physical interpretation, are stable for any non-zero value of the extra integration constant. When the integrating constant that deforms Schwarzschild is null, obviously, the rescaling and the constant conformal factor are not well defined, but in that case we remain only with the Schwarzschild black hole. We thank an anonymous PRD referee for requesting clarification about that.}. It deforms both the asymptotic behaviour of the metric and the geometry of the event horizon. In fact, the fall-off of the metric for large radial distance is not Minkowskian, and the shape of the event horizon is not perfectly spherical, as can be appreciated by picture (\ref{fig:picture-horizons}.b) and (\ref{fig:picture-horizons}.c). 
We can infer the distorting effect of the external gravitational field, determined by the parameter $B$, by analysing the equatorial  
\beq
        C_e \ = \ \int_0^{2\pi} \sqrt{g_{\varphi\varphi}} \; d\varphi \ = \  \frac{4m\pi}{1+m^2B^2}   \ ,
\eeq
and the polar
\beq
        C_p \ = \ 2 \int_{-1}^{1} \sqrt{g_{xx}} \; dx \ = \ 8m \frac{EllipticE(-m^2B^2)}{(1+m^2B^2)^2}
\eeq
circumferences\footnote{The $EllipticE$ function  is defined, as in Wolfram language, as $EllipticE(x) = \int_0^{\pi/2} \sqrt{1-x\sin^2\theta} \, d\theta$. $x\leq1$ in order to get real values.}. Since, for any $B\neq0$, $C_p<C_e<C_s$, where $C_s$ is the spherical Schwarzschild circumference (i.e. for $B=0$), we understand that the geometry of the black hole's event horizon gradually shrinks for growing absolute values of $B$. More specifically the polar circumference gets smaller faster with respect to the equatorial one, generating the oblate shape of the event horizon that can be recognised by the isometric embedding of the horizon into the Euclidean flat space, as drawn in figure \ref{fig:picture-horizons}. \\
The event horizon area
\beq \label{area}
        A = \int_0^{2\pi} d\varphi \int_{0}^{\pi} d\theta \sqrt{g_{\theta\theta} g_{\varphi\varphi}} \ = \ \frac{16 \pi m^2}{(1+B^2m^2)^3} \ ,
\eeq
is smaller with respect to the Schwarzschild one because the denominator of (\ref{area}) is always bigger than 1. \\

Considering small the extra parameter, which determines the curvature of the background, we can expand the metric (\ref{bh})-(\ref{Omega}), for instance up to the first significative order, to get
\bea \label{expansion}
       ds^2 &=& \left\{- \left(1-\frac{2m}{r} \right) + \ \left[m^2-m r + \frac{r^2}{2} + \ \left(-4m^2+5mr-\frac{3r^2}{2} \right) \cos^2\theta \right] B^2 + \mathcal{O}^3(B) \right\} dt^2 + \nn \\ 
       &+& \left\{ \frac{r}{r-2m} + \ \frac{r^2}{2(r-2m)^2} \left[2m^2+10mr-5r^2+(8m^2-10mr+3r^2)\cos^2\theta \right]\, B^2 \mathcal{O}^3(B) \right\}  dr^2 +  \nn \\
        &+&  \left\{ r^2 + \ \frac{r^2}{2} \left[-3r^2+(-2m^2-4mr+3r^2)\cos^2\theta \right]\, B^2+ \mathcal{O}^3(B) \right\} d\theta^2 + \\
        &+&  \left\{ r^2 \sin^2\theta + \ \frac{r^2}{2} \sin^2\theta \left[ -4m^2-r^2+(2m^2-4mr+r^2)\cos^2\theta  \right]\, B^2 + \mathcal{O}^3(B) \right\} d\varphi^2 \ . \nn
\eea
Obviously this metric is not an exact solution any more, but expanding to further orders we can get arbitrarily close to the exact solution (\ref{bh})-(\ref{Omega}). The expansion (\ref{expansion}) constitutes a small deviation from the Schwarzschild black hole, and it might represent a good description for possible gravitational deformation of the static spherically symmetric metric, at least not far from the black hole horizon. Moreover in this region, because of the small values of the parameter $B$, the deviation from the flat background can be made as small as desired.\\ 

\subsection{Thermodynamics} 

The temperature of the black hole defined from the surface gravity $\kappa_s$ is
\beq
       T = \frac{\kappa_s}{2\pi} = \frac{1}{2\pi} \sqrt{-\frac{1}{2} \nabla_\m \chi_\n \nabla^{\m} \chi^\n} =   \frac{(1+B^2m^2)^2}{8\pi m} \ .
\eeq
where $\chi=\p_t$ is the Killing vector generating the event horizon.\\
The mass of the black hole can be computed from the definition of the Komar integral
\beq \label{mass}
            M = \frac{1}{4\pi} \oint_{\S_\infty} \nabla^\m \chi^\n d\S_{\m\n} 
            = \frac{m}{1+B^2m^2} \ ,
\eeq
where the infinitesimal integration surface $d\S_{\m\n} = - n_{[\m}\s_{\n]} \sqrt{g_{\theta\theta}g_{\varphi\varphi}} \, d\theta d\varphi$ is integrated at spatial infinity; $n_\m$ and $\s_\n$ are two orthonormal timelike and spacelike vectors normal to the integrating surface $\S_\infty$.\\
Considering the Bekenstein-Hawking entropy for the black hole, that is a quarter of the event horizon area
\beq
              S = \frac{A}{4} \ ,
\eeq
we can verify the Smarr relation
\beq
               M = 2 T S \ .
\eeq
The first law of thermodynamics can be satisfied if the Killing vector $\chi^{\m}$ is properly normalised: $\chi^\m \to \bar{\chi}^\m = \chi^\m/\sqrt{1+m^2B^2}$. In this way the entropy remains unchanged, $\bar{S}=S$, but mass and the temperature rescale accordingly to become
\beq
           \bar{M} = \frac{m}{(1+m^2B^2)^{3/2}} \ , \hspace{2cm}   \bar{T} = \frac{(1+m^2B^2)^{3/2}}{8\pi m} \ .
\eeq
Then the first law is verified\footnote{The first law can be verified also in the canonical ensemble, that is in terms of the Helmholtz free energy $F$, which is defined as the Legendre transforming the mass: $\bar{F} = \bar{M} -\bar{T}\bar{S}=\bar{M}/2$. In fact $\d F=-\bar{S}\d \bar{T}$.}
\beq
            \d \bar{M} = \bar{T} \d \bar{S} \ ,
\eeq
without spoiling the Smarr relation for the barred quantities, indeed $\bar{M} = 2 \bar{T} \bar{S}$. Note that, with this normalisation of the time-like coordinate, also the Cristodolou-Ruffini formula \cite{christodoulou}, $\bar{M}^2=\bar{S}/(4\pi)$, is fulfilled.  This behaviour resembles much the thermodynamics of the Schwarzschild in the Bertotti-Robinson electromagnetic background \cite{kerr-bertotti}, with or without the extra Bonnor-Melvin electromagnetic field \cite{bh-brbm}. Also the position of the event horizon in (\ref{horizon}), coincides with the Schwarzschild-Bertotti-Robinson black hole.  However the interpretation of the integrating constant $B$ is different because, in the case presented here, it regards the intensity of the gravitational external back-reacting gravitational field, not the external Bertotti-Robinson electromagnetic field. \\
The local thermodynamic stability can be studied through the heat capacity $\bar{C}$ of the system, which is defined as the change of the heat $Q$ for a variation of the temperature or $\d\bar{Q}=\bar{C} \d \bar{T} = \bar{T} \d \bar{S}$. For the black hole here considered, observing that  $\bar{S}=1/(16\pi \bar{T}^2)$, we can compute the heat capacity
\beq
        \bar{C} \, = \, \bar{T} \,\frac{\p \bar{S}}{\p \bar{T}} \,  = \, - \frac{1}{8\pi\bar{T}^2} \, = \, -2\bar{S} \,  < \, 0 \ .
\eeq
Since the heat capacity is negative, the black hole is locally unstable from a thermodynamic point of view. Physically it means that it can evaporate or radiate, decreasing its mass when the temperature rises, which is mathematically expressed as follows
\beq
             \frac{\p \bar{M}}{\p \bar{T}} = - \frac{1}{8\pi\bar{T}^2} \ < 0 \ .
\eeq
This behaviour is in complete analogy with the Schwarzschild black hole. However we note that, when $B$ increases the heat capacity decreases (in absolute value), thus also the thermodynamic instability with respect to the Schwarzschild spacetime reduces. \\

\subsection{Background} 

The small mass limit can serve as a good approximation to describe the spacetime far from the black hole. In particular when we set $m=0$ in the main solution (\ref{bh})-(\ref{Omega}) we obtain the external gravitational background described by the metric
\beq \label{background}
      ds^2 = \left[\frac{1+\sqrt{1+B^2r^2\sin^2\theta}}{2\left( 1+B^2r^2\sin^2\theta \right)} \right]^2 \left[-(1+B^2r^2) \, dt^2 + \frac{dr^2}{1+B^2r^2} + r^2 d\theta^2 \right] + \frac{4r^2\sin^2\theta \ d\varphi^2}{\left(1+\sqrt{1+B^2r^2\sin^2\theta}\right)^2}  \ .  
\eeq
Notice that the background spacetime (\ref{background}) becomes of Petrov type D. In fact, following \cite{stephani-big-book}, we compute the Cartan scalar invariants and find $I^3=27J^2$, $I\neq N$ and $K=N=0$. This background is everywhere devoid of conical or curvature singularities; for instance the Kretschmann scalar invariant 
\beq
         R_{\m\n\s\l} R^{\m\n\s\l}  =   \frac{768 \, B^4 \, (1+B^2r^2\sin^2 \theta)^3\left[B^2r^2\sin^2\theta + 2 \left( 1 + \sqrt{1+B^2 r^2\sin^2 \theta}\right) \right]}{\left[1+\sqrt{1+B^2 r^2\sin^2 \theta} \right]^8}
\eeq 
is always bounded and thanks to (\ref{conicity}) it is possible to check that there are no irremediable conical singularities.  \\
The causal structure resembles the one of Anti de-Sitter, even though there are no cosmological constant terms in the field equations. That's because the Carter-Penrose diagram looks similar with respect to the negative constant curvature spacetime, where here the role of the cosmological constant is played by the parameter $B$. The main difference consists in the fact that the curvature is not constant (not even after a conformal transformation) because the base manifold is not a sphere as in $AdS$, but it is an oblate spheroid. When also the $B$ parameter becomes zero the background (\ref{background}) goes to the usual flat Minkowski space.\\
It is not an easy task to include the cosmological constant in the black hole solution (\ref{bh})-(\ref{Omega}) because the generating technique used to build that solution is not working well in the presence of the cosmological constant. In fact, the presence of the cosmological constant breaks some relevant symmetries of the Ernst equations \cite{charging}. Nevertheless, it is easy to find a cosmological extension of the background metric (\ref{background}), since it belongs to the special Petrov type D. It is possible to proceed by direct integration or casting the metric (\ref{background}) in a notable form. At this latter purpose we can perform the following change of coordinates
\beq
             r \to \frac{\sqrt{4p -4p^2+B^2q^2}}{B(2p-1)} \ , \qquad \theta \to \arccos \left( \frac{Bq}{\sqrt{4p -4p^2+B^2q^2}} \right) \ , \qquad \varphi \to \frac{B^2}{2} \, \varphi  \ ,
\eeq
to put the background (\ref{background}) into a subcase of the non-expanding type D solution of Kundt's class\footnote{Note that two different branches for the above transformation are possible, for $p>1/2$ and $p<1/2$. Here we follow explicitly the former case, which keeps positive the radial coordinate $r$, in any case, the latter is similar. When $p>1/2$ the curvature singularity of the metric (\ref{16.27}) at $p=0$ is not concerning.} as in eq. (16.27) of \cite{Griffiths-Podolsky}
\beq \label{16.27}
        ds^2 =  p^2 \left[ - Q(q) dt^2 + \frac{dq^2}{Q(q)} + \frac{dp^2}{P(p)} \right] + \frac{P(p)}{p^2} d\varphi^2          
\eeq
with 
\beq \label{QP}
          Q(q) =  1 + B^2 q^2 \ , \qquad \qquad P(p) = B^2 p - B^2 p^2 \ .
\eeq
Notice that this wide class of metrics includes also the Witten's bubble of nothing \cite{witten}, the  Bonnor-Melvin \cite{Griffiths-Podolsky} and the swirling  \cite{swirling} spacetimes. \\
Having reduced the background (\ref{background}) in the form of eq. (\ref{16.27}), it is well known how to extend it in the presence of the cosmological constant, just upgrading the $P(p)$ function in 
\beq
        P(p) = B^2 p - B^2 p^2 - \frac{\L}{3} p^4 \ .
\eeq
Furthermore the background written in the form of (\ref{16.27}) opens to an unexpected physical interpretation. In fact the metric (\ref{16.27}) - (\ref{QP}) is the double Wick rotation of the hyperbolic Schwarzschild metric. To prove this is sufficient to consider the topological generalisation of the Schwarzschild spacetime
\beq \label{topo-schwarzschild}
      d\hat{s}^2 =   - \left(\hat{k}- \frac{2\hat{m}}{\hat{r}} \right) \, d\hat{t}^2 + \frac{d\hat{r}^2}{\hat{k}- \frac{2\hat{m}}{\hat{r}}} + \frac{\hat{r}^2 \, d\hat{x}^2}{1-\hat{k}\hat{x}^2} + \hat{r}^2 (1-\hat{k}\hat{x}^2) \, d\hat{\varphi}^2 \ .
\eeq
The $\hat{k}$ parameter determines the curvature of the angular section of the metric, for constant time and radial coordinate ($\hat{t},\hat{r}$). This surface has constant curvature determined by the sign of $\hat{k}$, as pointed out by its Ricci scalar $R=2\hat{k}/\hat{r}^2$. Actually the Schwarzschild black hole is clearly retrieved from (\ref{topo-schwarzschild}) for $\hat{k}=1$, where the event horizon, located at $\hat{r}=2\hat{m}$, acquires a spherical shape. For zero or negative $\hat{k}$ the constant $(\hat{t},\hat{r})$ sections describe a flat or a hyperbolic constant curvature surface respectively\footnote{In these cases the black hole interpretation is lost because there is no event horizon, at least without cosmological constant. On the other hand with the inclusion of a negative cosmological constant the black hole interpretation can be recovered also for non positive horizon curvature, but at the price of renouncing to the standard AdS global asymptotic.}.\\
Remarkably the background metric (\ref{16.27}) - (\ref{QP}) can be precisely recovered by the analytic continuation of the ($\hat{t},\hat{\varphi}$) coordinates and the following parameter and coordinate redefinition
\beq
          \hat{t} \to i \varphi \ , \qquad \hat{\varphi} \to i t \ , \qquad \hat{k} \to -B^2 \ , \qquad \hat{m} \to - \frac{B^2}{2} \ , \qquad \hat{x} \to q \ , \qquad \hat{r} \to p \ .
\eeq
On the other hand, for positive $\hat{k}$, the double Wick rotation of the ($\hat{t},\hat{\varphi}$) coordinates applied to the topological Schwarzschild metric (\ref{topo-schwarzschild}) give the Witten's expanding bubble of nothing \cite{witten}, \cite{horowitz-bubble-baths}. Therefore the background can be considered a topological generalisation of the metric which describes the expanding bubble of nothing. However note that the name {\it topological bubble} or {\it hyperbolic bubble}  might be considered an abuse of notation or just a mathematical definition; in fact, from a physical point of view, the bubble interpretation for positive $\hat{k}$ is not viable because there is no horizon for the bubble if $\hat{k}\leq 0$. The absence of a Cauchy horizon in the background spacetime means that the time-like and radial coordinates do not change their nature; hence the black hole solution (\ref{bh})-(\ref{Omega}) can be regarded as static, in the domain of outer communication. This is contrary to the black hole inside the Witten's bubble, which is an expanding solution, more details in the section \ref{sec:bh+bubble}.  \\ 
Nevertheless the background spacetime, physically representing the external gravitational field where the black hole of section \ref{sec:geometry} is embedded, is a hyperbolic\footnote{Note that the hyperbolicity lays in the base manifold of the metric (\ref{topo-schwarzschild}) before the double Wick rotation, therefore it reflects on the actual $(t,q)$ section of the metric (\ref{16.27}).} variant of the bubble of nothing.\\

\subsection{Geodesics} 
\label{sec:geo}

Thanks to the Killing vectors inherited from the axisymmetric and stationary spacetimes, i.e. $\xi=\p_t$ and $\eta=\p_\varphi$, we can study, similarly to \cite{kerr-bertotti}, the geodesic motion of a massive uncharged test particle, of unitary rest mass $m_0$, in the black hole metric (\ref{bh})-(\ref{Omega}). The two Killing vectors assure that the energy $E_0$ and the angular momentum $L_0$ of the test particle is preserved. We consider the motion on the equatorial plane, for $\cos \theta = \pi/2$, then
\bea \label{E0}
              E_0 &=& - g_{\m\n} u^\m \xi^\n \ = \  \left(1- \frac{2m}{r} - m^2 B^2 \right) \, \frac{\left(1+\sqrt{1+B^2r^2} \right)^2}{4 \, (1+B^2r^2)} \, \dot{t} \ , \\
              L_0 &=& g_{\m\n} u^\m \eta^\n \ = \ \frac{4 \, r^2 \D_\varphi^2}{\left(1+\sqrt{1+B^2r^2} \right)^2} \, \dot{\varphi} \ , \label{L0}
\eea
where the dot represents the derivative with respect to the affine parameter.  From the geodesic equation, the metric compatibility and (\ref{E0})-(\ref{L0}), we have that the quantity
\bea
          \e \ = \ - g_{\m\n} u^\m u^\n  &=& \frac{-\left(1+\sqrt{1+B^2r^2} \right)^2 \, r \, \dot{r}^2}{4(1+B^2r^2)^3(r -2m-m^2B^2 r)}  + \\
          &+& \frac{4 E_0^2 r (1+B^2r^2)}{(r -2m-m^2B^2 r)\left(1+\sqrt{1+B^2r^2} \right)^2} - \frac{\left(1+\sqrt{1+B^2r^2} \right)^2 L_0^2}{4r^2\D_\varphi^2} \nn  
\eea 
is constant. Then we can express the geodesic equation as the equation for a classical particle of energy $\varepsilon_0 = E_0^2/2
$ under the effect of a potential $V(r)$ 
\beq
              \mezzo \,  \dot{r}^2 = \frac{16(1+B^2 r^2)^4}{(2+B^2r^2+2\sqrt{1+B^2r^2})^2} \left[\varepsilon_0 - V(r) \right]
\eeq
with 
\beq
          V(r) \, = \, \frac{1-\frac{2m}{r}-B^2m^2}{8(1+B^2r^2)}\left(1+\sqrt{1+B^2r^2} \right)^2\left[L_0^2\, \frac{2+B^2r^2+2\sqrt{1+B^2r^2}}{4r^2\D_\varphi^2} + \e \right]
\eeq
We can find the stable circular orbits looking for the minimum of the potential $V(r)$. The innermost stable circular orbits are obtained by the conditions $V'(r)= V''(r)=0$. These conditions can be solved analytically for $r_{ISCO}$ and $L_{ISCO}$, however the expressions are quite lengthy. Therefore we evaluate these exact expression for some numerical values of the physical parameters describing the mass and intensity of the external gravitational field. For instance, for $m=1$, $B=1/100$ and $\e=1$ we have a couple of results
\beq
             \{ r_{ISCO} = 6.094 \ , \ L_{ISCO}^2= 11.889 \} \ , \hspace{2cm} \{ r_{ISCO} = 16.679 \ , \ L_{ISCO}^2= 16.103 \} \ .
\eeq
We observe that the smaller radius of the circular orbit is slightly larger with respect both the Schwarzschild and Schwarzschild-Bertotti-Robinson black holes, for which $r_{ISCO}=3 r_h$. Moreover there is a second possibility for a radius hosting stable circular orbits. These effects can be ascribed to the presence of the external gravitational field. \\

\subsection{A new look at the black hole inside the expanding bubble of nothing} 
\label{sec:bh+bubble}

It's interesting to note that the metric (\ref{bh})-(\ref{Omega}) admits an imaginary rotation of the hair parameter $B \to i \bar{B}$, without modifying its Lorentzian signature, nor becoming complex.
\beq  \label{bh-dwr}
         ds^2 = \frac{(1-\bar{B}^2mr\cos^2\theta +\Omega)^2}{4\Omega^4} \left[- f(r) dt^2 + \frac{dr^2}{f(r)} + \frac{r^2 d\theta^2}{1-\bar{B}^2m^2\cos^2\theta} \right] + \frac{4 r^2 \sin^2 \theta \, (1-\bar{B}^2m^2\cos^2\theta) \, \D_\varphi^2 \, d\varphi^2}{(1-\bar{B}^2mr\cos^2\theta +\Omega)^2} \, ,
\eeq
where 
\bea
        f(r) &=& \left(1-\frac{2m}{r} + \bar{B}^2m^2\right)(1-\bar{B}^2r^2) \ , \\
        \Omega(r,\theta) &=& \sqrt{1-\bar{B}^2r[r+(2m-\bar{B}^2m^2r-r)\cos^2\theta]} \ .\label{Omega-dwr}
\eea
Clearly the single black hole interpretation immersed into an external gravitational field is spoiled, the casual structure is modified because the appearance of extra Killing and conformal horizons. \\
We note that for $\bar{B}=0$ we clearly get, from (\ref{bh-dwr})-(\ref{Omega-dwr}), the Schwarzschild black hole, while for $m=0$ we remain with the complex continuation (in the parameter $B$) of the background in (\ref{background}), that is
\beq \label{background-dwr}
      ds^2 = \left[\frac{1+\sqrt{1-\bar{B}^2r^2\sin^2\theta}}{2\left( 1-\bar{B}^2r^2\sin^2\theta \right)} \right]^2 \left[-(1-\bar{B}^2r^2) \, dt^2 + \frac{dr^2}{1-\bar{B}^2r^2} + r^2 d\theta^2 \right] + \frac{4r^2\sin^2\theta \ d\varphi^2}{\left(1+\sqrt{1-\bar{B}^2r^2\sin^2\theta}\right)^2}  \ .  
\eeq
Thanks to the change of coordinates
\beq
        r \to \frac{2\bar{m}\sqrt{\bar{r}^2-2\bar{m} \bar{r}+\bar{m}^2 \cos^2 \bar{\theta}}}{\bar{r} - \bar{m}} \ , \qquad \theta \to \arccos \left( \frac{\bar{m} \cos \bar{\theta}}{\sqrt{\bar{r}^2-2\bar{m} \bar{r}+\bar{m}^2 \cos^2 \bar{\theta}}} \right) \ , \qquad t \to 2\bar{m}  \bar{t} \ , \qquad \varphi \to \frac{\bar{\varphi}}{4\bar{m}} \ ,  
\eeq
and redefinition of the parameters
\beq
             \bar{B} \to \frac{1}{2\bar{m}} \ \ , \qquad \qquad \D_\varphi \to 1 \ ,
\eeq
we get exactly the double Wick rotated Schwarzschild
\beq \label{bubble-shw}
         ds^2 = - \, \bar{r}^2 \sin^2 \bar{\theta} \, d\bar{t}^2 \, + \, \frac{\bar{r} \, d\bar{r}^2}{\bar{r}-2\bar{m}} \,  + \, \bar{r}^2 d\bar{\theta}^2 \, + \, \left( 1 - \frac{2\bar{m}}{\bar{r}} \right) d\bar{\varphi}^2 \ ,
\eeq
which is the spacetime discovered by Witten and representing an expanding bubble of nothing \cite{witten}, \cite{horowitz-bubble-baths}, whose three-dimensional constant curvature surface is defined in (\ref{bubble-shw}) by $r=r_b=2\bar{m}$. The whole four-dimensional spacetime can be interpreted as the geometry in between two very large black holes \cite{bubble}. \\
Hence the interpretation of the metric (\ref{bh-dwr})-(\ref{Omega-dwr}) is clearly a Schwarzschild black hole inside an expanding bubble. The Petrov type is $I$, as the one of section \ref{sec:solution}. This solution was first built, using Weyl cylindrical coordinates ($t,\rho, z, \varphi$) in \cite{bubble}, see also \cite{kerr+bubble} for an explicit representation.
However we are able, here, to present this spacetime in spherical coordinates, which  makes the solution more intelligible.  \\
A change of coordinate that verifies how the full metric (\ref{bh-dwr})-(\ref{Omega-dwr}) can be mapped into the black hole in the bubble written in Weyl form, can be found in \cite{kerr+bubble}.\\
From (\ref{bh-dwr})-(\ref{Omega-dwr}) it stems that the event horizon is located at 
\beq
          r_+ = \frac{2 m}{1+\bar{B}^2 m^2} \ ,
\eeq
while the bubble horizon is for
\beq
 r_b = \frac{1}{\bar{B}} \ .
\eeq
In case we want the bubble horizon outside the black hole horizon, the following condition on the parameters must be fulfilled 
\beq \label{condition}
|\bar{B}|<1/m \ ;
\eeq  
on the contrary the bubble would be inside the black hole horizon. However this latter configuration can spoil the black hole interpretation, because in the domain of outer communication there may be present curvatures singularities. In fact apart from $r=0$ the Kretschmann scalar invariant can diverge where
\beq
   1- \bar{B}^2 m r x^2 \sqrt{1-\bar{B}^2 r \big[ r+(2 m -r -\bar{B}^2 m^2 r)x^2 \big]}             = 0 \ .            
\eeq  
 \\
When the condition (\ref{condition}) is fulfilled the loci of these possible divergences are hidden, because they lay beyond both the event and the bubble horizons. \\
Possible conical singularities of the metric (\ref{bh-dwr})-(\ref{Omega-dwr}), along the symmetry axis defined by $\theta=0$ and $\theta=\pi$, are again removable, in this case by setting 
\beq \label{Dphi-drw}
                \D_\varphi = \frac{1}{1 - \bar{B}^2m^2} \ .
\eeq 
In fact, computing the ratio between the length of the circumference and the radius of a small circle around the north and south pole of the two dimensional surface, section of constant time and radial coordinates of metric (\ref{bh-dwr}), we get 
\beq \label{conicity}
       \lim_{\theta \to 0} \left( \frac{\int_0^{2\pi} \sqrt{g_{\varphi\varphi}} \ d\varphi}{\theta \ \sqrt{g_{\theta\theta}}} \right)  \ = \  2 \pi \ = \ \lim_{\theta \to \pi} \left( \frac{\int_0^{2\pi} \sqrt{g_{\varphi\varphi}} \ d\varphi}{(\pi - \theta) \ \sqrt{g_{\theta\theta}}} \right) \ ;
\eeq
which means that the value of $\D_\varphi$ in (\ref{Dphi-drw}) ensure no conicity in the spacetime.\\

A swirling extension of the black hole in the bubble can be found in \cite{bh-brbm}. On the other hand the immersion of the black hole in the bubble into an external electromagnetic filed, while easily obtainable also from the analytical continuation of the parameter $B$ from the general solution of \cite{bh-brbm}, has to be scrutinized with care, in particular about the energy conditions of the electromagnetic field strength. \\


\newpage

\section{Conclusions}

In this article we present a new static black hole in four-dimensional vacuum general relativity, remarkably extending the Schwarzschild solution by the presence of an extra integrating constant ($B$), which deforms the geometry of both the black hole and the asymptotic of the background. In fact, the event horizon is not spherically symmetric, but oblate, while the background ceases to be asymptotically flat. These two features of the metric allow the solution to avoid no-hair or Jebsen-Birkhoff's theorems, which state that the Schwarzschild solution is the only static and spherically symmetric in the Einstein theory.\\
Thanks to the new integrating constant the metric belongs to a more general type with respect to the Schwarzschild solution, that is, according to the Petrov classification, $I$ instead of $D$.\\  
Since the extra integrating constant can be considered as small as one needs, the new black hole is a continuous deformation of the standard Schwarzschild solution, and therefore can be directly tested by astrophysical experiments and observations (photon spheres, shadows, gravitational lensing) to determine upper bounds to the hair value, which we presume to be small, compared to the mass of the black hole; surely smaller than the black hole mass theoretical constraint imposed to have a well defined event horizon.\\
The fact that the new black hole is a continuos deformation of the Schwarzschild solution and not a completely different and disconnected metric, is, in a certain sense, reminiscent of the uniqueness black hole theorems. The Schwarzschild metric just adapts to the deformation of the external background, but it is not completely denaturalised. So, according to this interpretation, the principles of uniqueness theorems might still be valid, but in a generalised sense, taking into account non-trivial asymptotic conditions. \\  

Furthermore this discovery opens to possible other generalisations of the Schwarzschild black hole or maybe even different kinds of static black holes, when the asymptotic is not Minkowskian. In particular the closer generalisation would be embedding the Schwarzschild black hole in the last unexplored background case suggested by the double wick rotation of (\ref{topo-schwarzschild}). In fact, while for positive and negative $\hat{k}$ the metric has been obtained and it describes a Schwarzschild-like black hole in the bubble or in its hyperbolic variant presented in \ref{sec:solution}, the $\hat{k}=0$ case remains open. Also electrovacuum, accelerating and rotating extensions might be achieved, following the same generating technique exploited to produce this uncharged and static version \cite{bh-brbm}. Some of these cases, because more general, could be even more phenomenologically interesting. The embedding of this black hole into external rotating backgrounds or in electromagnetic external backgrounds, both the Bertotti-Robinson and the Bonnor-Melvin, is considered in \cite{bh-brbm}. As future perspective, for describing actual astrophysical models, it would be interesting to extend the pure vacuum scenario and include in this picture also the contribution of the accretion disk's matter, as detected in most black hole observations.\\

\paragraph{Acknowledgements}
{\small I would like to thank Hideki Maeda, Marcello Ortaggio and Fiorenza de Micheli for valuable comments on this subject.
A Mathematica notebook containing the main solution presented in this article can be found as an ancillary file in the arXiv source folder or as supplementary material in the PRD journal.}\\
\vspace{0.3cm}




\newpage





\begin{thebibliography}{99}

\bibitem{schwarzschild}
K.~Schwarzschild,
{\it ``On the gravitational field of a mass point according to Einstein's theory''},
Sitzungsber. Preuss. Akad. Wiss. Berlin (Math. Phys. ) \textbf{1916} (1916), 189-196; 
\href{https://arxiv.org/pdf/physics/9905030}{\tt [arXiv:physics/9905030 [physics]]}

\bibitem{levi-civita-c}
T.Levi-Civita, {\it ``$ds^2$ einsteiniani in campi newtoniani''}, \href{http://operedigitali.lincei.it/rendicontiFMN/rol/pdf/S5V27T2A1918P343_351.pdf}{Atti Accad.Naz. Lincei Rend., \textbf{27}, 343 (1918)}.

\bibitem{chandrasekhar}
S.~Chandrasekhar,
{\it ``The mathematical theory of black holes''}, Clarendon Press, Oxford (1985)

\bibitem{bubble}
 M.~Astorino, R.~Emparan and A.~Vigan\`o,
 {\it ``Bubbles of nothing in binary black holes and black rings, and viceversa''},
  \href{https://doi.org/10.1007/JHEP07(2022)007}{JHEP \textbf{07} (2022), 007};
   \href{https://arxiv.org/pdf/2204.09690.pdf}{\tt [arXiv:2204.09690 [hep-th]]}.


\bibitem{swirling}
 M.~Astorino, R.~Martelli and A.~Vigan\`o,
  {\it ``Black holes in a swirling universe''},
   \href{https://doi.org/10.1103/PhysRevD.106.064014}{Phys. Rev. D \textbf{106} (2022) no.6, 064014};
    \href{https://arxiv.org/pdf/2205.13548}{\tt [arXiv:2205.13548 [gr-qc]]}


\bibitem{jebsen}
J.~T.~Jebsen,
{\it ``On the general spherically symmetric solutions of Einstein's gravitational equations in vacuo''}, Ark. Mat. Ast. Fys. (Stockholm) \textbf{15}, nr.18 (1921), reprint in \href{https://doi.org/10.1007/s10714-005-0168-y}{Gen. Rel. Grav. \textbf{37} (2005) no.12, 2253-2259} 


\bibitem{birkhoff}
G.D. Birkhoff, {\it ``Relativity and Modern Physics''}, p.253, Harvard University Press,
Cambridge (1923).

\bibitem{israel}
W.~Israel,
{\it ``Event horizons in static vacuum space-times''}, \href{https://doi.org/10.1103/PhysRev.164.1776}{Phys. Rev. \textbf{164} (1967), 1776-1779}

\bibitem{bh-brbm}
M.~Astorino,
{\it ``Black holes in the external Bertotti-Robinson-Bonnor-Melvin electromagnetic field''}, 
\href{https://doi.org/10.1103/c5lw-53yd}{Phys. Rev. D \textbf{112} (2025) no.10, 104077};
\href{https://arxiv.org/pdf/2508.12908}{\tt [arXiv:2508.12908 [gr-qc]]}

\bibitem{ernst1}
   F.~J.~Ernst,
  {\it ``New formulation of the axially symmetric gravitational field problem''},
  \href{https://doi.org/10.1103/PhysRev.167.1175}{Phys.\ Rev.\  {\bf 167} (1968) 1175}

\bibitem{ernst-magnetic} 
  F.~J.~Ernst,
   {\it ``Black holes in a magnetic universe''},
    \href{https://doi.org/10.1063/1.522781}{J.\ Math.\ Phys.\  {\bf 17}, no. 1, 54 (1976).}

\bibitem{stephani-big-book}
  H.~Stephani, D.~Kramer, M.~A.~H.~MacCallum, C.~Hoenselaers and E.~Herlt,
  {``Exact solutions of Einstein's field equations''}, \ 
  \href{https://doi.org/10.1017/CBO9780511535185}{\tt [doi:10.1017/CBO9780511535185]}

\bibitem{Griffiths-Podolsky}
J.~B.~Griffiths and J.~Podolsky,
{\it ``Exact Space-Times in Einstein's General Relativity''}, 
\href{https://doi.org/10.1017/CBO9780511635397}{Cambridge University Press, 2009}

\bibitem{Type-I}
M.~Astorino,
 {\it ``Accelerating and Charged Type I Black Holes''},
  \href{https://doi.org/10.1103/PhysRevD.108.124025}{Phys. Rev. D \textbf{108} (2023) no.12, 124025};
   \href{https://arxiv.org/pdf/2307.10534.pdf}{\tt [arXiv:2307.10534 [gr-qc]]}

\bibitem{christodoulou}
D.~Christodoulou and R.~Ruffini,
{\it ``Reversible transformations of a charged black hole''},
\href{https://doi.org/10.1103/PhysRevD.4.3552}{Phys. Rev. D \textbf{4} (1971), 3552-3555}


\bibitem{kerr-bertotti}
J.~Podolsky and H.~Ovcharenko,
{\it ``Kerr black hole in a uniform magnetic field: An exact solution''}, 
\href{https://doi.org/10.1103/rfgv-ybz5}{Phys. Rev. Lett. \textbf{135} (2025) no.18, 181401}; 
\href{https://arxiv.org/pdf/2507.05199}{\tt [arXiv:2507.05199 [gr-qc]]}

\bibitem{charging}
M.~Astorino,
{\it ``Charging axisymmetric space-times with cosmological constant''},
\href{https://doi.org/10.1007/JHEP06(2012)086}{JHEP \textbf{06} (2012), 086} ;
\href{https://arxiv.org/pdf/1205.6998.pdf}{\tt [arXiv:1205.6998 [gr-qc]]}


\bibitem{witten}
E.~Witten,
{\it ``Instability of the Kaluza-Klein Vacuum''},
\href{https://doi.org/10.1016/0550-3213(82)90007-4}{Nucl. Phys. B \textbf{195} (1982), 481-492}

\bibitem{horowitz-bubble-baths}
O.~Aharony, M.~Fabinger, G.~T.~Horowitz and E.~Silverstein,
{\it ``Clean time dependent string backgrounds from bubble baths''}, \href{https://doi.org/10.1088/1126-6708/2002/07/007}{
JHEP \textbf{07} (2002), 007};
\href{https://arxiv.org/abs/hep-th/0204158}{\tt [hep-th/0204158]}.

\bibitem{kerr+bubble}
M.~Astorino,
{\it ``Kerr Black Holes in an Expanding Bubble''},
\href{https://arxiv.org/pdf/2507.22114.pdf}{\tt [arXiv:2507.22114 [gr-qc]]} 

\end{thebibliography}
\end{document}